\documentclass[12pt,preprint]{aastex62} 
\usepackage{booktabs}
\def\lea{\mathrel{<\kern-1.0em\lower0.9ex\hbox{$\sim$}}}
\def\gea{\mathrel{>\kern-1.0em\lower0.9ex\hbox{$\sim$}}}

\graphicspath{{./}{./Figures/}{figures/}}
\received{}
\revised{}
\accepted{}
\submitjournal{ApJ}

\shorttitle{Constraints on Upper Cutoffs in the Mass Functions of Young Star Clusters}
\shortauthors{Mok et al.}

\begin{document}

\title{Constraints on Upper Cutoffs in the Mass Functions of Young Star Clusters}
\correspondingauthor{Angus Mok}
\email{mok.angus@gmail.com}
\author[0000-0001-7413-7534]{Angus Mok}
\affil{Department of Physics \& Astronomy, The University of Toledo, Toledo, OH 43606, USA}
\author{Rupali Chandar}
\affil{Department of Physics \& Astronomy, The University of Toledo, Toledo, OH 43606, USA}
\author{S. Michael Fall}
\affil{Space Telescope Science Institute, Baltimore, MD 21218, USA}

\begin{abstract}
We test claims that the power-law mass functions of young star clusters (ages $\lea\mbox{few}\times10^8$~yr) have physical upper cutoffs at $M_*\sim10^5~M_{\odot}$. Specifically, we perform maximum-likelihood fits of the Schechter function, $\psi(M)=dN/dM\propto M^{\beta}~\mbox{exp}(-M/M_*)$, to the observed cluster masses in eight well-studied galaxies (LMC, SMC, NGC 4214, NGC 4449, M83, M51, Antennae, and NGC 3256). In most cases, we find that a wide range of cutoff mass is permitted ($10^5~M_\odot \lesssim M_* < \infty$).  We find a weak detection at $M_* \sim 10^5~M_\odot$ in one case (M51) and strong evidence against this value in two cases. However, when we include realistic errors in cluster masses in our analysis, the constraints on $M_*$ become weaker and there are no significant detections (even for M51). Our data are generally consistent with much larger cutoffs, at $M_*\sim\mbox{few}\times10^6~M_{\odot}$. This is the predicted cutoff from dynamical models in which old globular clusters and young clusters observed today formed by similar physical processes with similar initial mass functions.
\end{abstract}

\section{Introduction} \label{sec:intro}
\par
One of the most important characteristics of a population of astronomical objects is its mass function, $\psi(M) = dN/dM$. The shape of this function, and especially any distinct features, such as upper or lower cutoffs, encodes important information about the physical processes involved in the formation and subsequent evolution of the objects. For young star clusters in different galaxies, the mass function is always found, in a first approximation, to have a power-law shape, $\psi(M) \propto M^{\beta}$, with an exponent close to $\beta \approx -2$, over the range from below $\sim 10^4~M_\odot$ to above $\sim10^5~M_{\odot}$ \citep[e.g.][]{Zhang99, Hunter03, Fall12, Krumholz15, Linden17}. Of course, to keep the total mass of clusters in a galaxy finite, mass functions with $\beta \approx -2$ must have both upper and lower cutoffs. The lower cutoff likely lies near the transition between individual stars and clusters of stars at $\sim10^2~M_{\odot}$ \citep[e.g.][]{Krumholz17}. The upper cutoff is the subject of this paper.
\par
As is customary, we represent the mass function of young star clusters by the  \citet{Schechter76} function, $\psi(M) = (\psi_*/M_*)~ (M/M_*)^{\beta} \exp(-M/M_*$), i.e., a power law with an exponent $\beta$ and an exponential cutoff at $M \approx M_*$. \citet{Fall01} introduced the Schechter mass function into this field in a theoretical study of the long-term disruption of star clusters. Their models match the observed mass function of globular clusters (with peaks at $M_p \sim10^5~M_{\odot}$) after $\sim10^{10}$ yr of evolution if the initial mass function has almost any shape, including a power law, and an exponential cutoff at $M_* \sim$ few $\times~10^6~M_\odot$ \citep[meaning $10^6 ~ M_\odot \lesssim M_* \lesssim 10^7 ~ M_\odot$; see also][]{Chandar07, Jordan07, McLaughlin08, Goudfrooij16}. 
\par
This prompted a search for upper cutoffs in the observed mass functions of recently formed clusters. Several such studies have claimed to detect cutoffs near $M_* \sim 10^5~M_\odot$ \citep[e.g.][]{Gieles09, Larsen09, Portegies10, Adamo15, Messa18}, far below the cutoff predicted by the \citet{Fall01} models. However, these claimed detections do not appear convincing by eye and have not been confirmed by robust statistical tests. The tests that have been performed are based on binned data and/or cumulative distributions. The purpose of this paper is to remedy this situation by performing maximum-likelihood fits of the Schechter function to the mass data for young clusters in eight well-studied galaxies, including some spirals and irregulars (LMC, M83, and M51) where cutoffs at $M_*\sim 10^5~M_{\odot}$ have been claimed \citep[e.g.][]{Larsen09, Adamo15, Messa18}. 
\par
The remainder of this paper is organized as follows. In Section 2, we describe the cluster samples, mass estimates, and mass distributions we use in this study. In Section 3, we describe the likelihood analysis we use to derive best-fit values and confidence contours for the parameters $\beta$ and $M_*$ in the Schechter function. We summarize our results and discuss their implications in Section 4.
\section{Cluster Mass Functions} \label{sec:data}
\begin{figure}[ht!]
\centering
\includegraphics[width = 10.5cm]{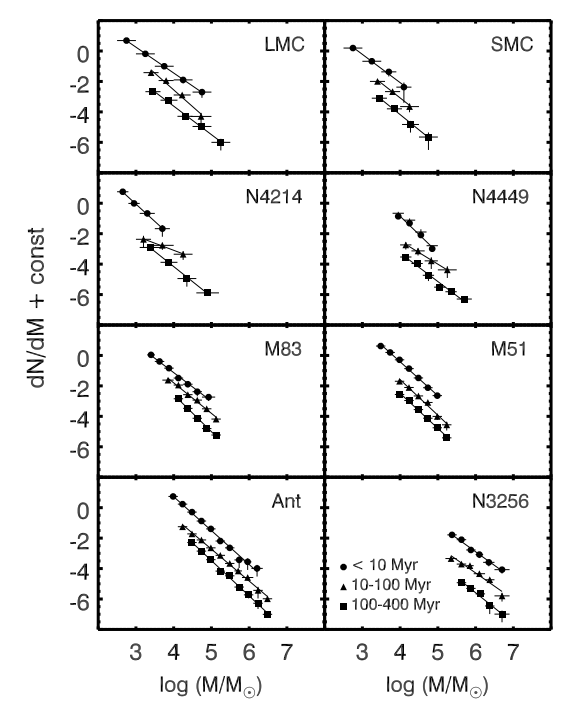}
\caption{Cluster mass functions with equal logarithmic bins in three age intervals, $\tau<10$ Myr (circles), $\tau=10-100$ Myr (triangles), and $\tau=100-400$ Myr (squares) for the 8 galaxies in our sample (as indicated). The straight lines show the best-fit power laws to each distribution, $\psi(M) \propto M^{\beta}$ \citep[reproduced from Figure 4 of ][]{Chandar17}. These binned mass functions are for visualization purposes only. All Schechter parameters are determined from the unbinned mass estimates of individual clusters using the maximum-likelihood analysis described in Section~3.}\label{fig:massfunc}
\end{figure}
\begin{figure}[ht!]
\centering
\includegraphics[width = 12.5cm]{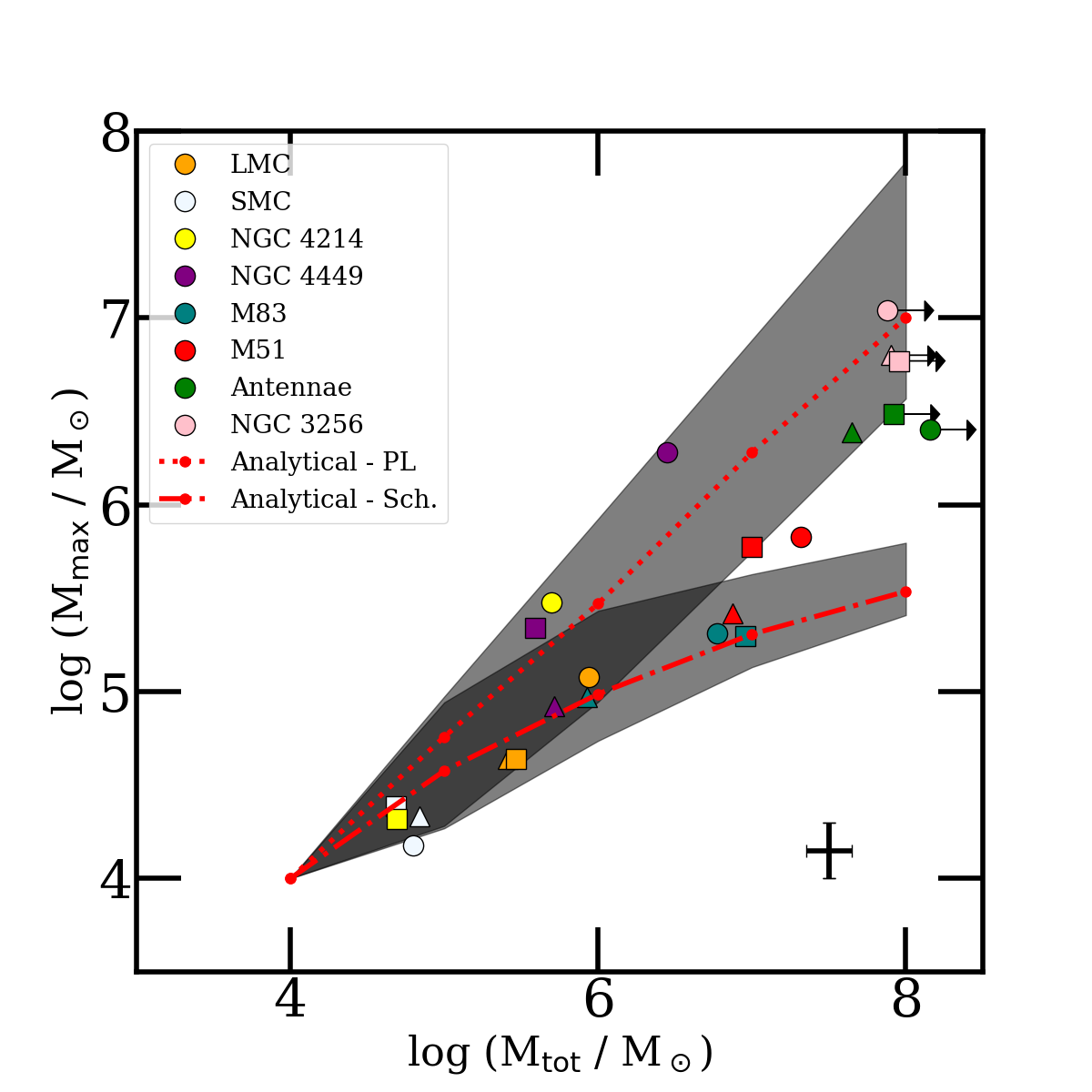}
\caption{Observed correlation between maximum cluster mass $M_{\rm max}$ and the total mass in clusters $M_{\rm tot}$ (with $M >10^4~M_\odot$) for the 8 galaxies in our sample. Triangles represent $<10$ Myr age bins, squares $10-100$ Myr bins, and circles $100-400$ Myr bins. Arrows indicate age bins with only lower limits on $M_{\rm tot}$, since their incompleteness limits are above $10^4~M_\odot$.   Uncertainties of 0.3 are shown in the lower-right, a typical value for $M_{\rm max}$ and an upper limit for $M_{\rm tot}$. The dotted and dot-dash red lines show the predicted statistical relations, respectively, for a pure power-law mass function with $\beta=-2$ and a Schechter mass function with $\beta=-2$ and $M_*=10^5~M_{\odot}$. The gray bands show the corresponding 95\% confidence regions as derived by bootstrap sampling. Evidently, most of the data-points are consistent with sampling from a pure power-law mass function (with $M_* \rightarrow \infty$).}\label{fig:mostmass}
\end{figure}
\par
We re-examine the mass functions of the cluster population in eight galaxies from our previous studies \citep{Chandar15, Chandar17}. The selection and photometry of clusters is based on $UBVI$H$\alpha$ images taken with the {\em Hubble Space Telescope} ($HST$) for NGC 4214 \citep{Chandar17}, NGC 4449 \citep{Rangelov11}, M83 (Whitmore et al. in prep), M51 \citep{Chandar16}, the Antennae \citep{Whitmore10}, and NGC 3256 \citep{Mulia16}, and $UBVR$ images taken with the Michigan Curtis Schmidt telescope for the LMC and SMC \citep{Hunter03}. Clusters were selected to be compact, but no attempt was made to distinguish bound from unbound clusters based on their appearance. The number of clusters in each galaxy varies from a few hundred (e.g., NGC~4449 and NGC~4214) to many thousands (e.g., M83, M51, the Antennae). The mass and age of each cluster were estimated by comparing the observed shape of the spectral energy distribution with predictions from the \citet{Bruzual03} stellar population models, assuming a \citet{Chabrier03} stellar IMF, and a Milky Way extinction law \citep{Fitzpatrick99}. Details of the observations, data reduction, and the cluster catalogs can be found in \citet{Chandar17} and the references therein\footnote{The 8 cluster catalogs are available at: \\ http://photon.panet.utoledo.edu/owncloud/index.php/s/GZempSsP7pWYJO4}.
\par
The major uncertainty in the mass estimates of young clusters comes from uncertainty in the ages, through the age-dependent mass-to-light ratios from the stellar population models. Clusters with ages $\tau \lea 10$~Myr and $\tau \approx100-400$~Myr have typical uncertainties of $\sim0.3$ in ~log~$M$, corresponding to a factor of $\sim2$ in $M$ \citep[e.g.][]{Hunter03, Fall05, Chandar10, deGrijs06}. The uncertainties may be larger for clusters with ages in the interval $\tau=10-100$ Myr, where the stellar population models show loops in color-color space, potentially leading to non-unique age and hence mass estimates. Errors in the distance or assumed stellar IMF will not affect the {\em shape} of the cluster mass function, although they will affect the normalization.  Stochastic fluctuations in the luminosities and colors of clusters may affect determinations of the mass function below $\sim3\times10^3~M_{\odot}$, but not at the higher masses of interest here \citep{Fouesneau12, Krumholz15, Krumholz18}.
\par
As in our previous studies, we divide the clusters into three age intervals: $<10$~Myr, $10-100$~Myr, and $100-400$~Myr \citep{Chandar15, Chandar17}. The oldest age interval of $100-400$~Myr is best suited to characterizing the cluster mass functions, because it is well populated, has reliable mass estimates, and uniform completeness. In contrast, the middle age interval ($10-100$~Myr) has uncertain and non-unique mass estimates, as mentioned above, while the youngest age interval ($<10$~Myr) has potential incompleteness due to dust obscuration and crowding \citep{Chandar14}.
\par
Figure~\ref{fig:massfunc} shows the binned mass functions of the cluster populations in our 8 galaxies in the three age intervals $<10$, $10-100$, and $100-400$ Myr \citep[reproduced from][]{Chandar17}. All of these mass functions are well represented by power laws, $\psi(M)\propto M^{\beta}$ with $\beta\approx-2$, and have no obvious bends or breaks.
Thus, any upper cutoff must occur near or beyond the maximum observed mass $M_{\rm max}$ in each sample of clusters. This circumstance raises the question of whether $M_{\rm max}$ is determined by a physical cutoff, as in the Schechter function, or by a statistical cutoff linked to the sample size.
\par
Figure~2 shows the observed maximum mass ($M_{\rm max}$) plotted against the observed total mass ($M_{\rm tot}$) in clusters more massive than $10^4~M_{\odot}$ (a proxy for sample size) in each of the three age intervals and eight galaxies\footnote{We omit the $<10$~Myr age bin for NGC~4449 because it has no clusters with $M>10^4~M_{\odot}$.}. We have also plotted the predicted statistical relations between $M_{\rm max}$ and $M_{\rm tot}$ for a pure power law with $\beta=-2$ and a Schechter function with $\beta=-2$ and $M_*=10^5~M_{\odot}$ (red dotted and dot-dash lines, respectively). These were computed from the requirement that the expected number of clusters more massive than $M_{\rm max}$ be unity. For the pure power law with $\beta=-2$, the $M_{\rm max}$--$M_{\rm tot}$ relation takes the particularly simple form
\begin{equation}
M_{\rm tot} = M_{\rm max} \times \ln \Big(\frac{M_{\rm max}}{10^4 M_\odot}\Big).
\end{equation}
\noindent Using the methodology in Chapter 5.3 of \citet{Bevington03}, we generate mock cluster catalogs from a power-law mass function with $\beta=-2$ and a Schechter function with $\beta=-2$ and $M_*=10^5~M_{\odot}$. Drawing 1000 random realizations to match key values of $M_{\rm tot}$, we plot connected gray bands in Figure~2 showing the region encompassing 95\% of the samples. Overall, the observed correlation between $M_{\rm max}$ and $M_{\rm tot}$ for our sample appears to follow more closely the predicted relation for a pure power law than that for a Schechter function with $M_*\sim10^5~M_{\odot}$. 

\begin{figure}[ht!]
\centering
\includegraphics[width = 16.5cm]{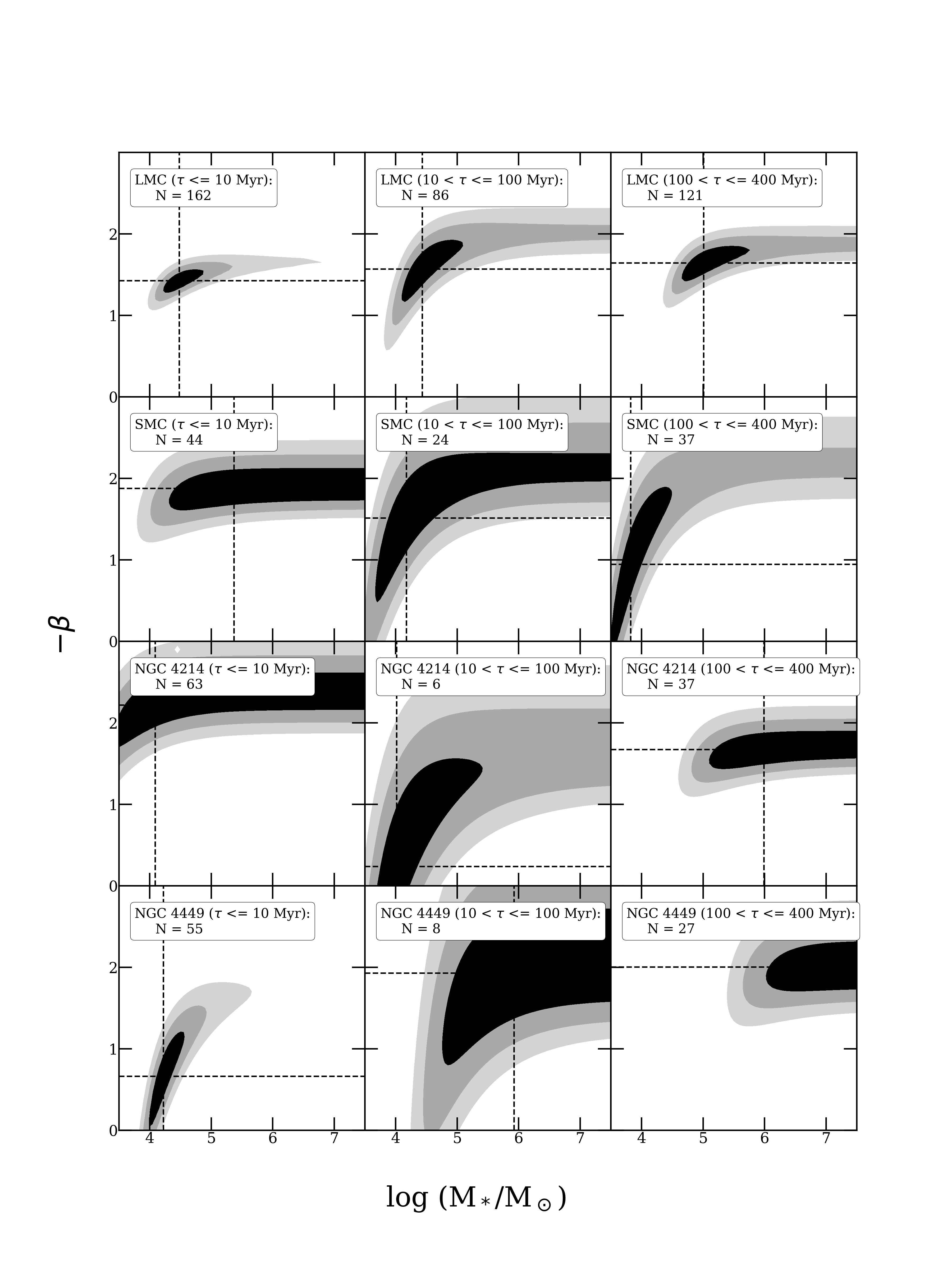}
\caption{Likelihood fits of the Schechter parameters $\beta$ and $M_*$ to the masses of clusters in the three age intervals $<10$~Myr (left), $10-100$~Myr (center), and $100-400$~Myr (right) in the four galaxies: LMC, SMC, NGC 4214, and NGC 4449. The dashed lines show the best-fit values of $\beta$ and $M_*$, while the boundaries of the shaded regions show the 1, 2, and 3$\sigma$ confidence contours.}\label{fig:likelm}
\end{figure}
\begin{figure}[ht!]
\centering
\includegraphics[width = 16.5cm]{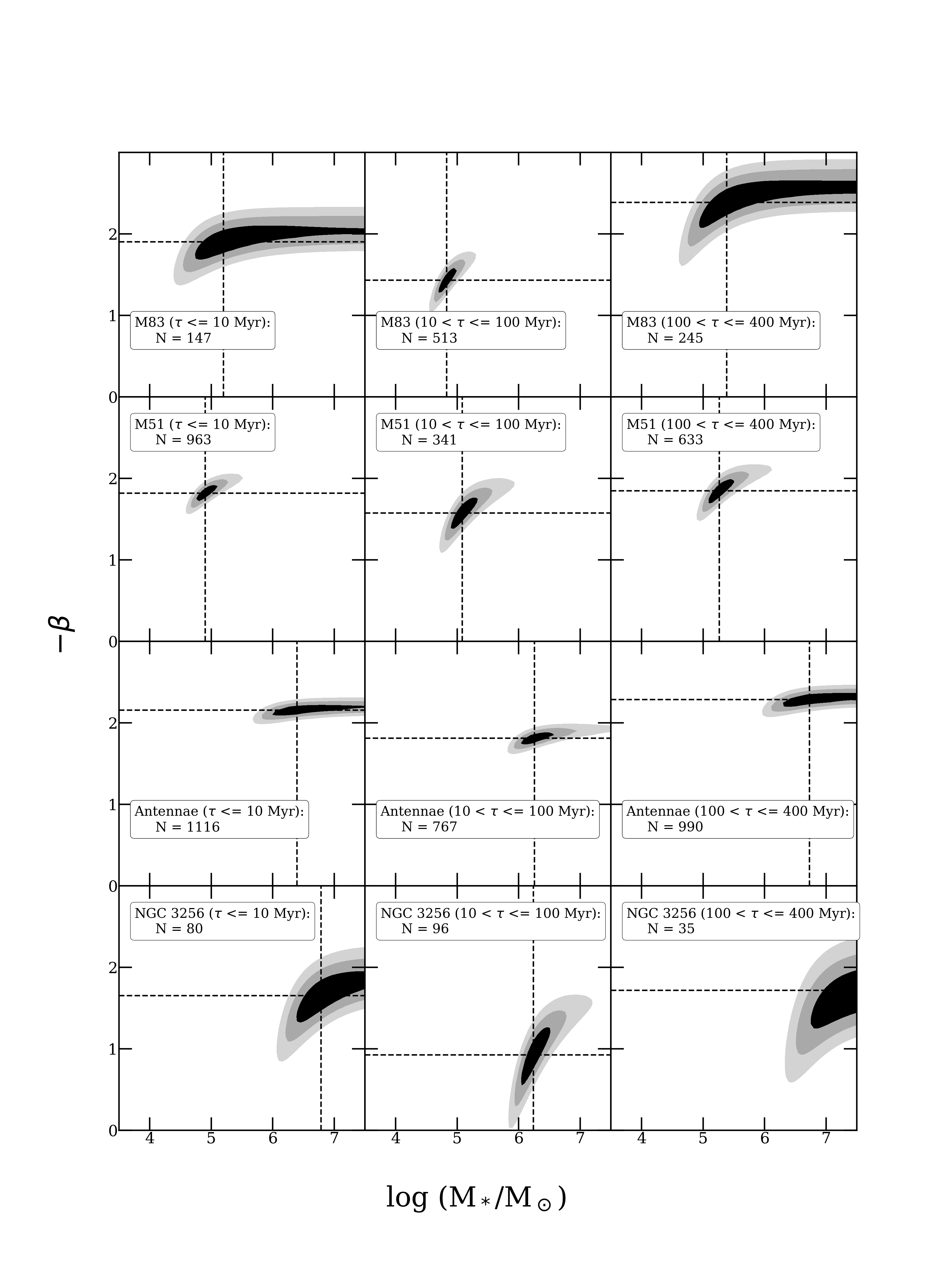}
\caption{Likelihood fits of the Schechter parameters $\beta$ and $M_*$ to the masses of clusters in the three age intervals $<10$~Myr (left), $10-100$~Myr (center), and $100-400$~Myr (right) in the four galaxies: M83, M51, the Antennae, and NGC~3256. The dashed lines show the best-fit values of $\beta$ and $M_*$, while the boundaries of the shaded regions show the 1, 2, and 3$\sigma$ confidence contours.}\label{fig:likehm}
\end{figure}

\section{Maximum Likelihood Fits} \label{sec:analysis}
\par
We now determine the best-fit values and confidence intervals of the parameters $\beta$ and $M_*$ in the Schechter function by the method of maximum likelihood. This method has the advantages of not requiring binned data (where weak features at the ends of the distribution may be hidden) or cumulative distributions (where the data points are not independent of one another). We follow the procedure described in detail in Chapter 15.2 of \citet{Mo10} for fitting a Schechter function to discrete luminosity or mass data. Specifically, we compute the likelihood $L(\beta,M_*)=\prod\limits_i P_i$ as a function of $\beta$ and $M_*$, where the probability $P_i$ for each cluster is given by
\begin{equation}
P_i = \frac{\psi(M_i)}{\int_{M_{\rm min}}^{M_{\rm max}} \psi(M) dM},
\end{equation}
\noindent and the product is over all clusters in the sample in question. We adopt the $M_{\rm min}$ values listed in Table 4 of \citet{Chandar17}, which stay above the completeness limit of each sample, and we set $M_{\rm max}=10^{7.5} M_\odot$ in all cases. We use the Nelder-Mead (1965) algorithm to find the maximum likelihood $L_{\rm max}$, and the standard formula
\begin{equation}
\ln L(\beta,M_*) = \ln L_{\rm max} - \frac{1}{2}\chi^2_p(k),
\end{equation}
\noindent where $\chi^2_p (k)$ is the chi-squared distribution with $k$ degrees of freedom and $p$ confidence level \citep{Mo10}, to derive the corresponding confidence contours. We have checked all our results using a Markov Chain Monte Carlo (MCMC) routine, and find similar contours whenever these close around best-fit values of $\beta$ and $M_*$. For the cases of contours that extend to the right edge, the corresponding contours derived from the MCMC routine are slightly smaller than those derived from equation~(3), leading to tighter lower limits on $M_*$.
\par
Figures~\ref{fig:likelm} and \ref{fig:likehm} show the best-fit values of $\beta$ and $M_*$ (dashed lines), and the 1, 2, and $3\sigma$ confidence contours (shaded regions) derived from equation~(3) for each of the three age intervals, $<10$~Myr (left panels), $10-100$~Myr (middle panels), and $100-400$~Myr (right panels), and for each galaxy in our sample. The best-fit values of $\beta$ and $M_*$ and their $2\sigma$ uncertainties are listed in Table~1. The resulting shapes of the confidence contours partly reflect correlations between the exponent $\beta$ and cutoff $M_*$, since there is a trade-off between steeper $\beta$ and larger $M_*$, and vice versa.
\par
\begin{figure}[ht!]
\centering
\includegraphics[width = 16.5cm]{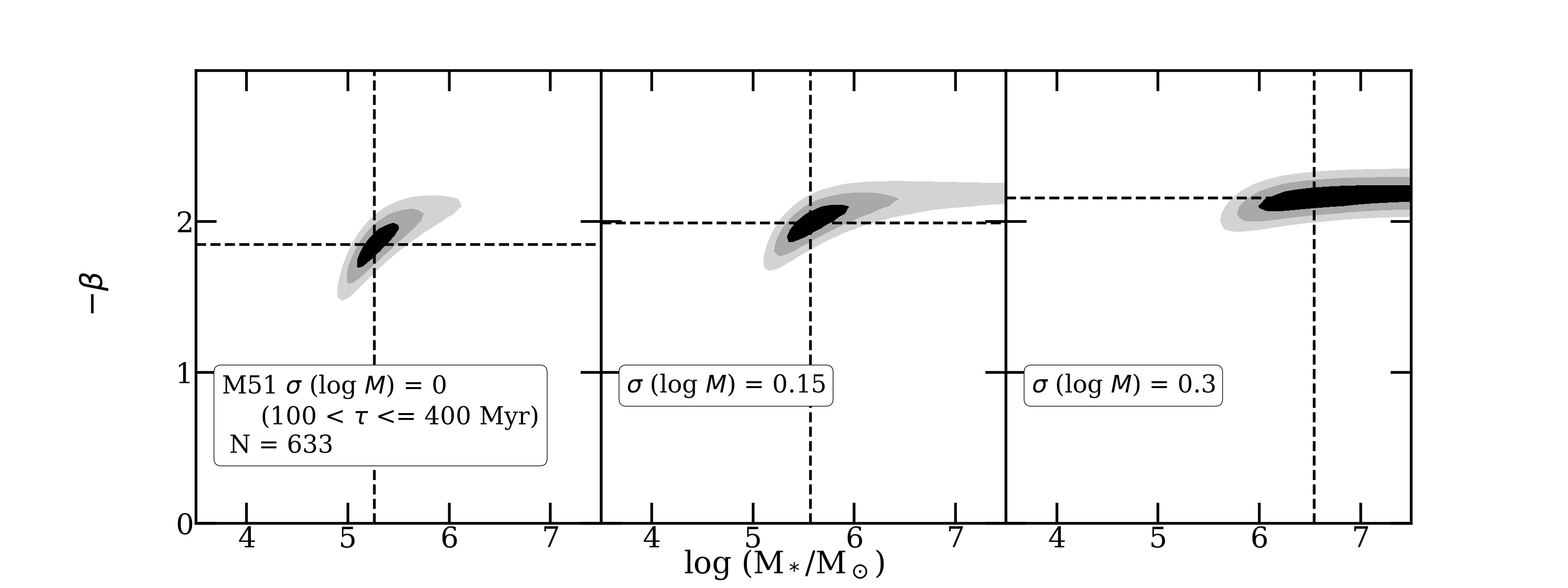}
\caption{Likelihood fits of the Schechter parameters $\beta$ and $M_*$ to the masses of clusters in M51 for the $100-400$~Myr age interval (left), after convolving with a Gaussian error distribution with a standard deviation of 0.15 dex (center) and 0.3 dex (right) (the latter being the most likely uncertainty). The dashed lines show the best-fit values of $\beta$ and $M_*$, while the boundaries of the shaded regions show the 1, 2, and 3$\sigma$ confidence contours. Note that the statistical significance of $M_*\sim10^5~M_\odot$ disappears as $\sigma(\log M)$ increases from 0 to 0.3.}\label{fig:errorcomp}
\end{figure}

Thus far, we have neglected uncertainties in the mass estimates of the clusters. The uncertainties in the Schechter parameters $\beta$ and $M_*$ shown in Figures~\ref{fig:likelm} and \ref{fig:likehm} and listed in Table~1 are therefore actually lower limits to the true uncertainties. To determine how much uncertainties in the mass estimates affect the fitted parameters, we convolve the Schechter mass function with a log-normal error distribution (Gaussian in log $M$), with a standard deviation $\sigma (\log M)$ \citep[see][]{Efstathiou88, Ratcliffe98}. We adopt $\sigma (\log M) = 0$ (as before), 0.15, and 0.30 (the typical uncertainties in individual mass estimates discussed in Section 2). In all cases, we find that the best-fit value of $M_*$ increases and its statistical significance decreases, with increasing $\sigma (\log M)$. As an example, we show the results for the $100-400$ Myr age bin in M51 in Figure~\ref{fig:errorcomp}. Evidently, the possible detection of an upper cutoff at $M_*\sim10^5 M_\odot$ for $\sigma (\log M) = 0$ disappears for $\sigma (\log M) = 0.3$. 

\begin{figure}[ht!]
\centering
\includegraphics[width = 16.5cm]{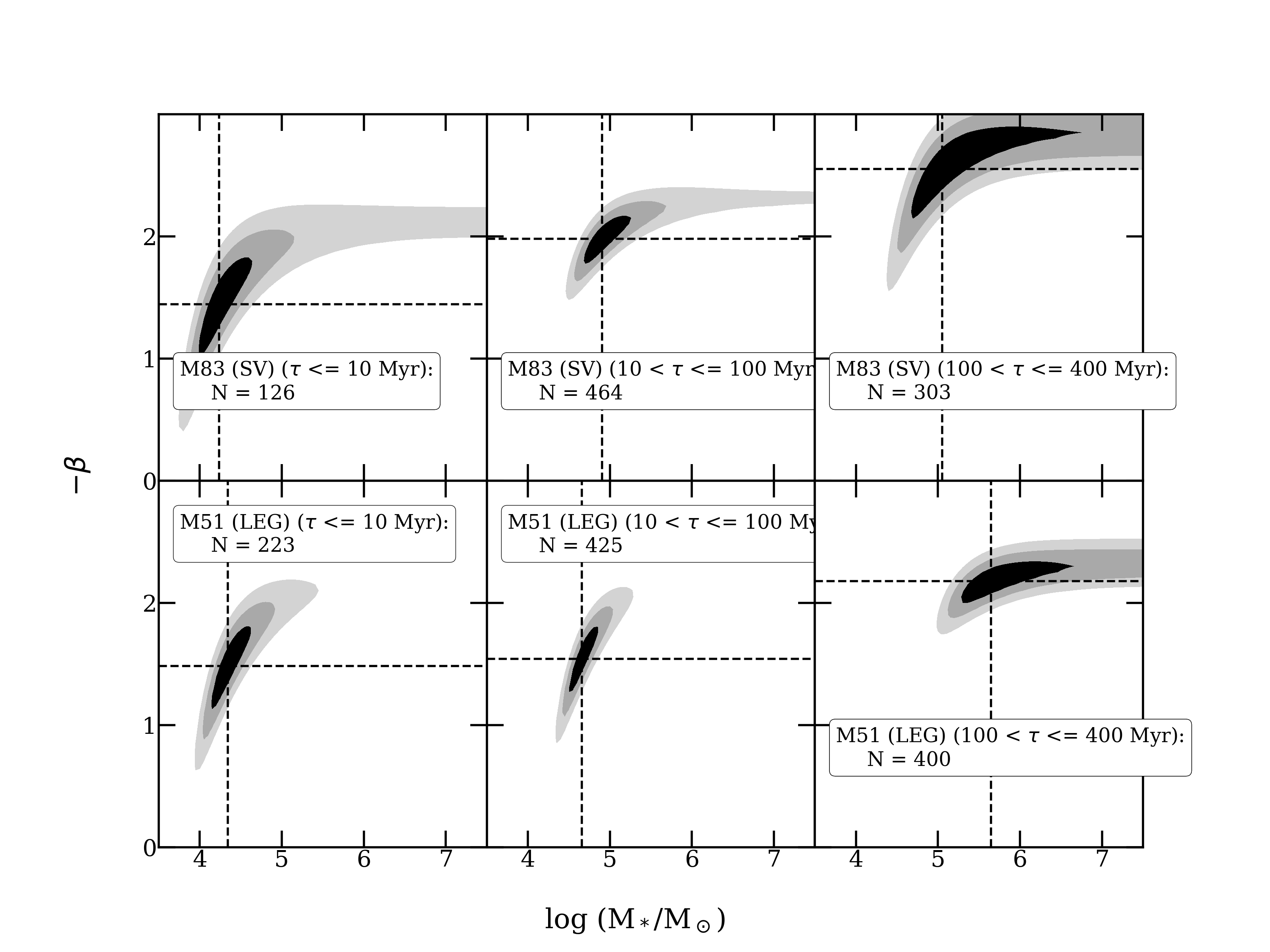}
\caption{Likelihood fits of the Schechter parameters $\beta$ and $M_*$ to the masses of clusters in the three age intervals $<10$~Myr (left), $10-100$~Myr (center), and $100-400$~Myr (right) in two galaxies with independent catalogs: M83 \citep{Silva-Villa14} and M51 \citep{Messa18}. The dashed lines show the best-fit values of $\beta$ and $M_*$, while the boundaries of the shaded regions show the 1, 2, and 3$\sigma$ confidence contours. Note that these confidence contours are generally similar to or less restrictive than those shown in Figure~\ref{fig:likehm} for our own catalogs of M51 and M83 clusters.}\label{fig:likecomp}
\end{figure}

Next, we check whether the fitted parameters $\beta$ and $M_*$ are robust with respect to different cluster catalogs by repeating our analysis with the \citet{Silva-Villa14} catalog of M83 clusters and the LEGUS catalog of M51 clusters \citep[see][]{Messa18}. The results of these tests are shown in Figure~\ref{fig:likecomp}. Compared to the confidence contours for our cluster catalogs, those for the other catalogs are generally similar or less restrictive. In particular, the $2\sigma$ confidence contour is unbounded for the $100-400$ Myr clusters in the LEGUS catalog of M51 clusters. 
\begin{deluxetable*}{ccccccc}[ht!]
\tablecaption{Best-Fit Schechter Parameters $\beta$ and $M_*$ and Their 2$\sigma$ Uncertainties\label{tab:galprop}}
\tablewidth{0pt}
\tablehead{
\colhead{ Galaxy} & \multicolumn{2}{c}{$<10$ Myr} & \multicolumn{2}{c}{$10-100$ Myr} & \multicolumn{2}{c}{$100-400$ Myr} \\
\cmidrule(r){2-3} \cmidrule(r){4-5} \cmidrule(r){6-7}
\colhead{} & \colhead{$-\beta$} & \colhead{log $M_*$} & \colhead{$-\beta$} & \colhead{log $M_*$} & \colhead{$-\beta$} & \colhead{log $M_*$} 
}
\startdata
LMC & 1.42 [1.15, 1.65] & 4.48 [4.10, 5.35] & 1.57 [0.85, 2.15] & 4.43 [3.95, 7.50] & 1.65 [1.25, 2.00] & 5.01 [4.50, 7.50] \\
SMC & 1.88 [1.40, 2.30] & 5.37 [4.00, 7.50] & 1.51 [0.00, 2.70] & 4.18 [3.50, 7.50] & 0.95 [0.00, 2.40] & 3.82 [3.50, 7.50] \\
NGC~4214 & 2.22 [1.45, 2.85] & 4.09 [3.50, 7.50] & 0.24 [0.00, 2.00] & 4.02 [3.55, 7.50] & 1.67 [1.25, 2.05] & 5.99 [4.80, 7.50] \\
NGC~4449 & 0.55 [0.00, 1.55] & 4.22 [3.90, 4.95] & 1.93 [0.00, 3.00] & 5.92 [4.45, 7.50] & 2.00 [1.50, 2.55] & $>7.5$ [5.65, 7.50] \\
M83 & 1.90 [1.55, 2.20] & 5.20 [4.55, 7.50] & 1.43 [1.15, 1.70] & 4.83 [4.60, 5.15] & 2.38 [1.85, 2.80] & 5.38 [4.75, 7.50] \\
M51 & 1.82 [1.65, 2.00] & 4.90 [4.65, 5.30] & 1.58 [1.25, 1.90] & 5.08 [4.80, 5.60] & 1.85 [1.60, 2.10] & 5.26 [5.00, 5.75] \\
Antennae & 2.16 [2.05, 2.25] & 6.40 [5.80, 7.50] & 1.81 [1.65, 1.95] & 6.26 [5.90, 6.95] & 2.29 [2.15, 2.40] & 6.73 [6.10, 7.50] \\
NGC~3256 & 1.65 [1.10, 2.10] & 6.79 [6.20, 7.50] & 0.92 [0.30, 1.50] & 6.24 [5.95, 6.80] & 1.72 [1.25, 1.95] & $>7.5$ [6.35, 7.50] \\
\enddata
\end{deluxetable*}

\section{Discussion and Conclusions} \label{sec:conclusion}
\par 
The main results of this paper are displayed graphically in Figures~\ref{fig:likelm} and \ref{fig:likehm}. Before discussing these results in detail, we offer a few general remarks. Ideally, we should find consistent estimates of, or limits on, the Schechter parameters $\beta$ and $M_*$ in all three age intervals, because we do not expect the physics of cluster formation to change significantly over the relatively short period spanned by our data ($~4\times~10^8$~yr, i.e. $\sim3\%$ of the Hubble age). Thus, it would be physically implausible for $M_*$ to increase with age, although it could in principle decrease as a result of the preferential disruption of the most massive clusters. Nevertheless, systematic errors potentially affect mass estimates and sample completeness differently in the three age intervals. The upper cutoff $M_*$ is particularly sensitive to the presence or absence of only a few clusters and any errors in their masses.
\par
As we have already noted, sample completeness is likely lowest in the youngest age interval ($<10$~Myr), due to dust obscuration and crowding, while systematic errors and non-uniqueness in mass estimates are likely highest in the middle age interval ($10-100$~Myr) due to loops in the color tracks of stellar population models. Moreover, the middle age interval also tends to have the smallest number of clusters and thus the largest sampling errors. This leaves the oldest age interval ($100-400$~Myr) as the most reliable for determining the parameters $\beta$ and $M_*$ in the Schechter mass function. This age interval is well populated with clusters, has a higher degree of completeness, and more reliable mass estimates.
\par
With these remarks in mind, we now group the results shown in Figures~\ref{fig:likelm} and \ref{fig:likehm} into three broad categories based largely on our likelihood analysis in the {\em oldest age interval}. The first category, which includes the LMC, SMC, NGC~4214, and M83 shows {\em no evidence for} a cutoff. For these galaxies, a wide range of $M_*$ is allowed by the long horizontal confidence contours that start below $10^5~M_{\odot}$ and continue without closing to the right edge of the diagrams at $10^{7.5}~M_{\odot}$. This means that the cluster masses are consistent with being drawn from a pure power law, but that an upper cutoff (over this mass range of $M_*$) cannot be ruled out. We note that the large allowable ranges in $M_*$ for the LMC, SMC, and NGC~4214 are driven in part by the relatively small number of clusters in these galaxies. 
\par
A second category, which includes NGC~4449, the Antennae, and NGC~3256, shows {\em evidence against} an upper mass cutoff near $M_*\sim10^5~M_{\odot}$. While the confidence contours for these galaxies also remain unbounded up to the maximum adopted value of $10^{7.5}~M_{\odot}$, they do not extend down to $10^5~M_{\odot}$. The youngest age interval in NGC~4449 has closed contours that suggest a value of $M_*\sim 10^4~M_{\odot}$, but this is inconsistent with the contours for the oldest age interval. This galaxy, in particular, violates the physical principle noted above that $M_*$ should not increase rapidly with age. 
\par
M51 is the only galaxy in our sample that shows evidence, at the $\sim3\sigma$ level, for a cutoff near $M_*\sim10^5~M_{\odot}$, when no uncertainties on cluster mass estimates are included. However, as shown in Figure~5, when realistic errors in mass estimates ($\sim0.3$~log~$M$) are included, no statistically significant cutoff near $10^5~M_{\odot}$ is found in M51 either.  It is worth noting that this cutoff could also be explained if only a few clusters were missing from the catalogs as a result of dust obscuration and/or crowding. Furthermore, in a sample of 8 galaxies, there is a non-negligible probability ($34/2.4$\%) that an upper cutoff will be detected with marginal significance ($2/3\sigma$) in one of them (e.g., M51) even if the underlying mass function of clusters is a pure power law.
\par
In conclusion, there are four galaxies in our sample (LMC, SMC, NGC 4214, and M83) for which a wide range of cutoff mass is permitted ($10^5~M_{\odot} \lesssim M_* < \infty$), one galaxy (NGC 4449) for which our analysis gives an unphysical result, and two galaxies (Antennae and NGC 3256) for which an upper mass cutoff at $10^5~M_{\odot}$ is excluded. Only for M51 is there a possible detection at $M_* \sim10^5~M_{\odot}$, but even this becomes insignificant when we include realistic errors in cluster masses in our analysis. On the other hand, much higher cutoffs, at $M_*\sim\mbox{few}\times10^{6}~M_{\odot}$, are consistent with our likelihood analysis in nearly all cases. The higher cutoffs are needed to reconcile the mass functions of young clusters observed today with those of old globular clusters, assuming they formed by similar physical processes with similar initial mass functions, as in the \citet{Fall01} models.

\acknowledgments
We thank Mark Krumholz, Paul Goudfrooij, and the anonymous referees for helpful comments. R. C. acknowledges support from NSF grant 1517819. Based on observations made with the NASA/ESA Hubble Space Telescope, obtained from the data archive at the Space Telescope Science Institute. STScI is operated by the Association of Universities for Research in Astronomy, Inc. under NASA contract NAS 5-26555. This work makes use of Python and Python packages, including {\sc numpy} and the optimization package from {\sc scipy} \citep{scipy} for the minimization routine.


\end{document}